# Efficient Steering Mechanism for Mobile Network-enabled UAVs

Hamed Hellaoui[1], Ali Chelli[2], Miloud Bagaa[1], and Tarik Taleb[1,3]

[1]Communications and Networking Department, Aalto University, Finland. Email:firstname.lastname@aalto.fi
[2]Faculty of Engineering and Science, University of Agder, 4898 Grimstad, Norway. Email: ali.chelli@uia.no
[3]University of Oulu, 90570 Oulu, Finland.

*Abstract*—The consideration of mobile networks as a communication infrastructure for unmanned aerial vehicles (UAVs) creates a new plethora of emerging services and opportunities. In particular, the availability of different mobile network operators (MNOs) can be exploited by the UAVs to steer connection to the MNO ensuring the best quality of experience (QoE). While the concept of traffic steering is more known at the network side, extending it to the device level would allow meeting the emerging requirements of today's applications. In this vein, an efficient steering solutions that take into account the nature and the characteristics of this new type of communication is highly needed. The authors introduce, in this paper, a mechanism for steering the connection in mobile network-enabled UAVs. The proposed solution considers a realistic communication model that accounts for most of the propagation phenomena experienced by wireless signals. Moreover, given the complexity of the related optimization problem, which is inherent from this realistic model, the authors propose a solution based on coalitional game. The goal is to form UAVs in coalitions around the MNOs, in a way to enhance their QoE. The conducted performance evaluations show the potential of using several MNOs to enhance the QoE for mobile network-enabled UAVs and prove the effectiveness of the proposed solution. *Index Terms*—Unmanned Aerial Vehicles (UAVs), Mobile Networks, Connection steering, Game theory.

## I. Introduction

Recent years have been marked by the widespread use of UAVs, also called drones. UAVs have a wide range of applications, such as smart monitoring, surveillance, and rescue missions. To increase the coverage of drones' related services, the next generation of UAVs is relying on mobile networks as a communication infrastructure. This will enable beyond visual line-of-sight applications and open a new plethora of emerging services. Mobile networks also provide new opportunities that can enhance the QoE and meet the requirements of today's applications. One advantage is to exploit the availability of different MNOs and to steer UAVs' traffic through the network ensuring the best QoE. To ensure the feasibility of the MNO steering scheme, the UAV can be equipped with multiple radio interfaces. Each radio interface allows the UAV to connect to one MNO.

The concept of traffic steering is more known at the network side. It has been used to direct the flow of traffic through different means. The traffic-steering concept can be extended to the device level allowing to enhance the QoE for these devices. However, such solutions must take into account the nature and the characteristics of mobile networkenabled UAVs. For instance, the close to free-space signal propagation of the flying UAVs is translated into increased interference on the non-serving BSs, in the uplink scenario. Real-field evaluations showed that flying UAVs experience different quality of service compared to user equipment on the ground [1]. The efficiency of a steering mechanism would relate to its ability to evaluate UAV's QoE and to select, for each one, the adequate MNO.

In the literature, few works have addressed the connection steering problem for mobile network-enabled UAVs. In addition, previous works on this topic did not consider adequate channel models for these emerging communications. This underpins the focus of this paper wherein the authors propose an efficient steering mechanism for mobile network-enabled UAVs. The main contributions of the paper are the following: • We consider a realistic communication model for mobile network-enabled UAVs taking into account the fast fading, path loss, and interference. Moreover, given the complexity of the related optimization problem, which is inherent from this realistic model, a solution based on coalitional game is proposed to select for each UAV the MNO ensuring the best QoE.

- We provide numerical results for connection steering using several mobile networks. The obtained results demonstrated the potential of steering the connection in enhancing the QoE for the flying UAVs. In addition, our analyses demonstrate the effectiveness of the proposed coalitional game in achieving optimal QoE compared to a random selection scheme.
- We show that as the number of MNOs increases, the aggregate network sum-payoff, as well as the individual payoff of the UAVs increase leading to a reduced outage and improved QoE.

The rest of this paper is organized as follows. Section II reviews some works related to connection steering in vehicular communications. The system model and the problem formulation are provided in Section III. The proposed solution for connection steering in mobile-enabled UAVs is thereafter introduced in Section IV. Performance evaluations are conducted and provided in Section V. Section VI concludes the paper.

## II. Related Work

The concept of steering the traffic is more employed at the network side to direct the traffic through different network functions in a way to meet the expected quality of service (QoS). This allows to deal with the growing network traffic demand and ensure a good QoS for the users [2].

This concept has also been extended to steer the traffic from wireless devices connected to the network. In [3], the authors proposed a connection steering protocol for LTE-connected vehicles. Their approach is based on anticipating the QoS/QoE degradation and exploiting the different radio access technologies (e.g. LTE, WIFI) to direct the traffic accordingly. QoS/QoE-aware policies are defined and communicated to the users to select the most adequate radio access out of the available ones.

The potential applications of UAVs have been explored in different studies and projects. For instance, in [4], the authors have studied the use of an LTE network for realizing downlink data and uplink control communication with flying UAVs. In [5], the authors proposed to shift part of the control communication to be performed by the UAVs, reducing therefore network degradation. Such network degradation caused by cellular UAVs has been studied in [6]. Other works, such as [7], proposed the use of UAVs as flying BSs to provide connectivity to ground devices.

However, most of prior works have studied scenarios, whereby only a single mobile network is considered in their communication models. The authors in [8] have proposed to steer the connection of UAVs among different mobile operators. This is performed based on the perceived signal strength (RSSI) from each network. However, RSSI is not considered as a good QoS indicator when it comes to flying UAVs. Real-field evaluations showed that the received signals in the air are generally very good in terms of strength, but are different in terms of quality. In this paper, we consider a realistic communication model to evaluate the QoE for cellular UAVs. The framework of coalitional game is used, to enable connection steering, which has the potential of achieving a fair usage of resources [9].

III. SYSTEM MODEL AND PROBLEM FORMULATION

This section aims to model the system and formulate the underlying problem. It first derives the system model for one cellular network. Thereafter, a consideration of many mobile networks is provided along with the problem formulation.

*A. System Model*

The *uplink scenario* is considered in which the data is transmitted from the UAVs to the BSs. Let U be the set of UAVs and B the set of BSs. Let us denote by $u \in U$ the transmitting UAV, and by $v \in B$ the receiving/serving BS. In the proposed model, we take into consideration the interference from the other UAVs on the serving BS $v$. Let us denote by $t$ the interfering UAV node. The received signal at the BS $v$ can be expressed as follows:

$$y_v = \alpha_{uv}\sqrt{P_u}x_u + \sum_{t \in U}^{t \neq u} \alpha_{tv}\sqrt{P_t}x_t + n_v, \quad (1)$$

where $P_u$ and $P_t$ denote the transmission powers at node $u$ and node $t$, and $x_u$ and $x_t$ refer to the transmitted symbols by those devices ($u$ and $t$), respectively. The fading coefficients of the links $uv$ and $tv$ are referred to as $\alpha_{uv}$ and $\alpha_{tv}$, respectively. The noise $n_v$ is modelled as a zero-mean additive white Gaussian process with variance $N_0$. Let $\gamma_{uv}$ stand for the instantaneous received signal-to-noise ratio (SNR) for the link $uv$, which can be expressed as

$$\gamma_{uv} = P_u \alpha_{uv}^2 / N_0. \quad (2)$$

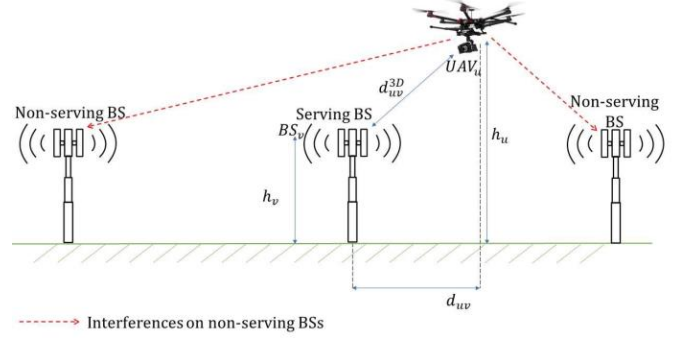

Fig. 1: System model (uplink).

*For the uplink*, the instantaneous received signal-to-interference-plus-noise ratio (SINR) for the link between a device $u$ and the BS $v$ can be obtained as

$$SINR_{uv} = \gamma_{uv}/(1 + \sum_{t \in U}^{t \neq u} \gamma_{tv}). \quad (3)$$

The propagation channel is modeled considering line-of-sight (LoS) and non line-of-sight (NLoS) links. We adopt the model proposed by 3GPP [1] to compute the probability of a LoS condition between a UAV $u$ and a BS $v$ as

$$P_{uv}^{LoS} = \begin{cases} 1 & \text{if } h_u > 100 \\ 1 & \text{if } d_{uv} \leq d_1 \\ \frac{d_1}{d_{uv}} + \exp\left(\frac{-d_{uv}}{p_1}\right)\left(1 - \frac{d_1}{d_{uv}}\right) & \text{if } d_{uv} > d_1, \end{cases} \quad (4)$$

with $d_1 = \max(460 \log_{10}(h_u) - 700, 18)$ and $p_1 = 4300 \log_{10}(h_u) - 3800$. The altitude of the UAV $u$ is denoted by $h_u$ and $d_{uv}$ is the 2D distance to the serving BS $v$, as shonw in Fig. 1. Note that the probability of an NLoS condition $P_{uv}^{NLoS}$, can be evaluated as $P_{uv}^{NLoS} = 1 - P_{uv}^{LoS}$. The path loss expression [1] depends on this condition and can be computed as

$$PL_{uv} = \begin{cases} 28.0 + 22\log_{10}(d_{uv}^{3D}) + 20\log_{10}(f_c) & \text{for LoS link} \\ -17.5 + (46 - 7\log_{10}(h_u))\log_{10}(d_{uv}^{3D}) + 20\log_{10}(\frac{40\pi f_c}{3}) & \text{for NLoS link} \end{cases} \quad (5)$$

where $f_c$ is the carrier frequency and $d_{uv}^{3D}$ is the Euclidean distance between the UAV $u$ and the BS $v$, as shown in Fig. 1. The effect of fast fading is taken into account in the proposed communication model. The fast fading follows a Nakagami-m distribution for LoS links, and a Rayleigh distribution for NLoS links. The mean values of the SNR for the LoS and NLoS links are referred to by the parameters $A_{uv}$ and $B_{uv}$, respectively, as follows

$$\begin{cases} A_{uv} = P_u/N_0 \times 10^{-\frac{PL_{uv}}{10}} = P_{uv}^{LoS} \\ B_{uv} = P_u/N_0 \times 10^{-\frac{PL_{uv}}{10}} = P_{uv}^{NLoS} \end{cases} \quad (6)$$

*Theorem 1:* In the uplink communication, a UAV $u$ fails in transmitting its packets to the BS $v \in \mathcal{B}$ iff $SINR_{uv}$ falls below a threshold $\gamma_{th}$. This event, called outage, occurs with a probability $P_{out,uv}^{UAV}(\gamma_{th})$ that can be expressed as

$$P_{out,uv}^{UAV}(\gamma_{th}) = \sum_{j=1}^{m} \beta_{1j} \frac{(-1)^j}{(j-1)!} \left(\frac{m}{A_{uv}}\right)^{-j} \Gamma(j) + \sum_{t=1}^{N} \delta_t' f_{j,1}(B_{tv}) - \sum_{t=1}^{N}\sum_{j=1}^{m} \delta_{t,j} \frac{(-1)^j}{(j-1)!} f_{j,j'}(A_{tv}/m) - \beta_{21} B_{uv} \left(1 + \exp\left(-\frac{\gamma_{th}}{B_{uv}}\right)\right) \left[\sum_{t=1}^{N} \frac{\delta_t'}{\frac{\gamma_{th}}{B_{uv}} + \frac{1}{B_{tv}}} - \sum_{t=1}^{N}\sum_{j=1}^{m} \frac{\delta_{t,j}}{\frac{\gamma_{th}}{B_{uv}} + \frac{m}{A_{tv}}} \frac{(-1)^j}{(j-1)!} \Gamma(j)\right] \quad (7)$$

where $\Gamma(j)$ is the gamma function. ([1,...,N]) refers to the list of interferer UAVs. The terms $\beta_{1j}$, $\beta_{21}$, $\delta_t'$, and $\delta_{t,j}$ have unique values satisfying the following formulas (fractional decomposition)

$$\left(1 - \frac{1}{A_{uv}}\right)^{-m} \left(1 - \frac{1}{B_{uv}}\right)^{-1} = \sum_{m} \beta_{1j} + (x - \beta_{21}) \times x \times B^{uv}$$

$$\sum_{j=1}^{m} \beta_{1j} \left(x - \frac{m}{A_{uv}}\right)^j \frac{1}{B_{uv}} \quad (8)$$

$$\prod_{t=1}^{N} (1 - xB_{tv})^{-1} (1 - xA_{tv})^{-m} = \sum_{t=1}^{N} \frac{\delta_t'}{1 - (x - A_{tv})} + \sum_{t=1}^{N}\sum_{j=1}^{m} \frac{\delta_{t,j}}{B_{tv}} \frac{1}{(x - \frac{m}{A_{tv}})^j} \quad (9)$$

The function $f_{j,j'}(S)$ is provided as

$$f_{j,j'}(S) = \sum_{p=1}^{n} S^j (\theta_p)^{j-1} \lambda_p \Gamma\left(j, \frac{m\gamma_{th}(\theta_p S + 1)}{A_{uv}}\right) \quad (10)$$

where $\lambda_p$ and $\theta_p$ denote respectively the weight and the zero factors of the $n$-th order Laguerre polynomials [10]. $\Gamma(a,z)$ is the upper incomplete gamma function defined as $\Gamma(a,z) = \int_z^{\infty} t^{a-1} e^{-t} dt$.

*Proof:* The proof is provided in [11]. ■

Theorem 1 provides the outage probability for mobile network-enabled UAVs on the uplink. It reflects the QoE a UAV would experience when transmitting data. The outage probability in Theorem 1 has been derived taking into account the path loss, fast fading, and interference. This makes the system model realistic since it considers most of the propagation phenomena that the wireless signal undergoes.

### B. Problem Formulation

To formulate the connection steering problem based on the above model, let us consider a set $\mathcal{O}$ of $O$ mobile operators providing connection through their deployed base stations. We assume that the MNOs can cooperate to ensure better QoE for the UAVs. Each UAV is connected to different mobile networks and transmits data through only one network at a given time. The goal is to steer the connection to the MNO ensuring the best QoE for each UAV. Let us denote by $uv_o$ the link between the UAV $u$ and its serving BS $v_o$ from the MNO $o \in \mathcal{O}$. The problem would therefore be translated into choosing for each UAV the serving BS from the available MNOs, while minimizing their outage probability.

To characterize the choice to be taken by each UAV, we define the Boolean variable $x_{uo}$ as follows

$$x_{uo} = \begin{cases} 1 & \text{If the UAV } u \text{ chooses the MNO } o \\ 0 & \text{Otherwise.} \end{cases} \quad (11)$$

Consequently, the steering problem would be expressed as

$$\min \max_{x} P_{outUAV,uvo}(\gamma_{th}) \quad (12)$$
$$u \in V, u_o$$

s.t.

$$\sum_{o \in O} u_o x = 1, \forall u \in U \quad (13)$$

$$x_{uo} \in \{0,1\}, \forall u \in U, \forall o \in O. \quad (14)$$

In the above optimization problem, constraint (13) ensures that each UAV selects one and only one MNO to be used for transmitting data. Constraint (14) limits the value of the decision variable to {0,1}. On the other hand, the objective function (12) aims to reduce the outage probability for the UAVs. This function is non convex and very complex. This complexity is inherent from the consideration of most of the propagation phenomena characterizing wireless communication. This shows the difficulty of achieving an optimal solution and raises the necessity of new solutions. The next section presents our proposed coalitional game optimization for connection steering in mobile network-enabled UAVs.

## IV. A COALITIONAL GAME FOR CONNECTION STEERING

In order to steer the connection to the MNO ensuring the best QoE for the UAVs, this paper proposes a coalitional game-based solution. A summary of the used notations is provided in Table I. The game is defined among the set of users V, where each UAV is considered as a player. The goal is to form the coalitions, such that the payoff of the different players is maximized.

A coalition $S_o$ represents a set of players that will rely on the MNO $o$ for communication (a UAV $u \in S_o$ will be served by its corresponding serving BS in the MNO $o$). Note that the number of coalitions equals to the number of available MNOs. Let $S = \{S_1, S_2, ..., S_O\}$ be the set of coalitions ($S_o \subseteq V$). As each UAV uses one MNO to transmit data at a given time, each two coalitions involve entirely different set of TABLE I: Summary of Notations.

| Notation | Description |
|---|---|
|  | Set of MNOs. $|O| = O$ |
|  | Set of players. |
| $S$ | Set of coalitions. $S = \{S_1, S_2, ..., S_O\}$ |
| $\Pi_{S_o}(u)$ | Payoff of the player $u \in S_o$ |
| $w(S_o)$ | Characteristic function of the coalition $S_o$ |
| $S_i \triangleleft_u S_j$ | The transfer function of the player $u$ from the coalition $S_i$ to the coalition $S_j$ |

players; i.e. $\forall S_1, S_2 \in S : S_1 \neq S_2 \Rightarrow S_1 \cap S_2 = \emptyset$. The payoff of each player $u$ belonging to a coalition $S_o$ can be obtained as follows

$$\Pi_{S_o}(u) = 1 - P_{outUAV,uvo} \left( \sum_{o \in O} \right), u \in S_o. \quad (15)$$

As we can see from (15), the payoff of a player is defined based on the outage probability of the corresponding UAV within the coalition. In fact, the payoff in (15) represents the probability of successful communication for player $u$. The player increases its payoff by reducing its outage probability and consequently increasing its success probability. As for the characteristic function of a coalition, it is defined as

$$w(S_o) = \sum_{u \in S_o} \Pi_{S_o}(u). \quad (16)$$

It is worth mentioning that the players are selfish and each one aims to increase its payoff without caring about the others. They change their coalitions in order to obtain a better payoff, leading therefore to decreased outage probability for all the players in their corresponding coalitions. To this end, we define the transfer operation which allows the UAVs to change their coalitions. This operation should ensure that the resulting partitioning is associated with a better payoff for the set of players.

*Definition 1: A player $u$ belonging to a coalition $S_i$ ($u \in S_i$) would be transferred to another coalition $S_j$ ($S_i \neq S_j$) iff:*

$$S_i \triangleleft_u S_j \Leftrightarrow \begin{array}{l} \Pi_{S_j \cup \{u\}}(u) > \Pi_{S_i}(u) \quad (17.1) \\ And \; w(S_i \setminus \{u\}) - w(S_i) > w(S_j) - w(S_j \cup \{u\}) \quad (17.2) \end{array} \quad (17)$$

The definition in (17) means that a player $u$ would be transferred from a coalition $S_i$ to another coalition $S_j$, if the concerned player will increase his payoff after the transfer (materialized by the condition (17.1)), while the gain of this operation on the coalition $S_i$ is larger than the loss on the coalition $S_j$ (condition (17.2)). Indeed, transferring a player from coalition $S_i$ to $S_j$ could potentially enhance the payoff of coalition $S_i$ (withdrawing a potential interferer) and decrease the payoff of coalition $S_j$. Condition (17.2) ensures that if the transfer operation incurs a loss on coalition $S_j$, this loss should not be larger than the benefit obtained by coalition $S_i$. The players keep changing their coalitions in order to enhance their payoffs.

Algorithm 1 illustrates our steering solution which is based on coalitional game. The execution of the game starts with an initial partition of the players on the coalitions. This initial partitioning is performed in a random manner. For each two coalitions $S_i, S_j \in S$, the transfer operation is Algorithm 1 Coalitional Game Algorithm.

Require: $S = \{S_1,...,S_O\}$
1: while True do
2:     Stable = True
3:     for each two coalitions $S_i, S_j \in S$ do
4:         for each player $u \in S_i$ do if $S_i \triangleleft_u S_j$
5:         then
6:             $S_i = S_i \setminus \{u\}$
7:             $S_j = S_j \cup \{u\}$
8:             Stable = False
9:         end if
10:        end for
11:    end for
12:    if Stable then
13:        break
14:    end if
15: end while

evaluated (lines [3 - 5] of Algorithm 1). This evaluation is performed according to equation (17). If the transfer conditions are satisfied, the two coalitions will be updated (lines [6 - 7] of Algorithm 1). This process will be repeated.

An important feature in coalitional game is the stability. A stable partition is reached if the players have no incentive to leave their coalitions since no player can increase his payoff by moving from one coalition to another. The stable partition is the optimal solution that maximizes the total sum-payoff. If no stable partition exists, the coalitional game is unstable. The variable 'Stable' in Algorithm 1 is used to characterize this state.

*Theorem 2: Starting from an initial random partition of the players on the coalitions, Algorithm 1 is guaranteed to converge towards a final stable and optimal partition.*

*Proof:*

As defined in Algorithm 1, the initial partition will be subject to players transfers applied sequentially. Let us express this transfer as follows:

$$S^{(0)} \to S^{(1)} \to S^{(2)} \to ... \quad (18)$$

where each $S^{(i)}$ represents the set of coalitions, $S$, after transfer operation number $i$ and $S^{(0)}$ is the first partition. The symbol

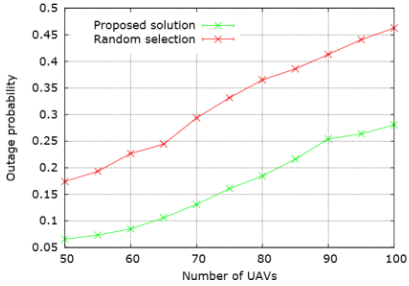
(a) Two MNOs

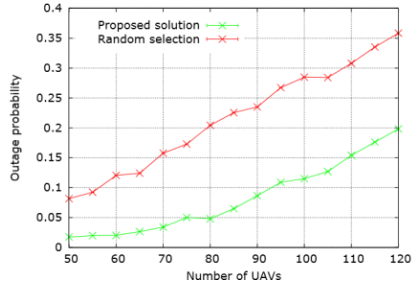
(b) Three MNOs

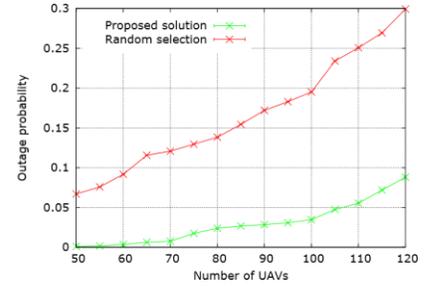
(c) Four MNOs

Fig. 2: The performance evaluation of the proposed coalitional game scheme for varying number of UAVs and MNOs.

$\to$ reflects a transformation operation from one state to another which is materialized by a transfer of one player. As the number of coalitions and the number of players are limited, the possible states of the coalitions are also limited. *Lemma 1: To prove the convergence of Algorithm 1, it suffices to prove that the transfer operation does not lead to repeated partitions.*

The above lemma is justified by the fact that the number of partitions is limited. If the partitions do not repeat, the sequence defined in (18) will converge to a final partition. This sequence is governed by the transfer operation defined in (17). The latter can also be written as follows

$$S_i \triangleleft_u S_j \Leftrightarrow \begin{cases} \Pi_{S_j \cup \{u\}}(u) > \Pi_{S_i}(u) \\ \text{And} \\ w(S_i \setminus \{u\}) + w(S_j \cup \{u\}) > w(S_j) + w(S_i). \quad (19.2) \end{cases} \quad (19)$$

From condition (19.2) we can see that the resulting states of the two concerned coalitions, together, have better payoffs compared to their original states. In addition, the other coalitions, not concerned by the transfer operation, will not be affected. In other words, their payoffs remain the same. Consequently, we can write the following

$$S^{(i)} \to S^{(j)} \Rightarrow \sum_{S_k \in S^{(i)}} w(S_k^{(i)}) < \sum_{S_l \in S^{(j)}} w(S_l^{(j)}). \quad (20)$$

Consequently, the transfer operation leads to different partitions. As per Lemma 1, Algorithm 1 does not lead to repeated partition and therefore converges to a final stable partition. Moreover, the sum of the payoffs of the resulting coalitions, after a transfer operation, increases as illustrated by (20). This shows that the final obtained partition has the largest sum-payoff and is thus optimal, which proves Theorem 2. ∎

## V. Performance Evaluation

In this section, we present the evaluation results of the proposed coalitional game for connection steering in mobile network-enabled UAVs. The communication model is implemented considering a Nakagami model with $m = 2$, a carrier frequency $f_c$ of 2 GHz, a noise variance $N_0$ of $-130 dBm$ [12], and a sensitivity threshold $\gamma_{th}$ of $10^{-3}$ [11]. The evaluation is performed in 1 km x 1 km square area. The altitude of the UAVs is randomly attributed between 22.5 m and 300 m, which is the applicability range for the used path loss model [1]. In each evaluation, we have used 12 BSs per MNO, with varying number of UAVs and MNOs.

Fig. 2 depicts the benefit of our coalitional game based solution, on the outage probability, for varying number of UAVs and MNOs. Our proposed scheme is compared to a random selection of the serving MNO by each UAV. The different sub-figures, Fig. 2(a), Fig. 2(b) and Fig. 2(c), have been obtained considering respectively two, three and four MNOs. These curves resulted from averaging the outage probability of all the considered UAVs. As we can see from these sub-figures, for a fixed number of UAVs, increasing the number of the MNOs reduces the outage probability for these UAVs. Indeed, as each UAV selects only one MNO to be used for transmitting data, the other non-serving MNOs will not be subject of interference from this UAV. We can also see that the outage probability is reduced, when increasing the number of MNOs, even with a random selection. This shows the potential of exploiting several mobile networks to enhance the QoE for flying UAVs. In addition, Fig. 2 illustrates the effectiveness of the proposed solution in enhancing the QoE for the flying UAVs. The MNO selection based on the coalitional game achieves better outage probability compared to the random selection, for different numbers of MNO and UAV. The coalitional game starts with a random selection, on which a sequence of player transfer operations will be applied. As shown in equation (20), the transfer operation enhances the sum of the characteristic function of the coalitions. Consequently, the final selection provided by the game ensures better payoff for the players, which is translated into reduced outage probability for the corresponding UAVs.

In Fig. 3, we have evaluated the sum of the payoffs for a fixed number of UAVs (120 UAVs), and different number of MNOs. The sum of the payoffs also reflects the sum of the coalitions' characteristic functions. As it can be seen from this figure, the sum-payoff increases with the number of considered MNOs. Since we consider a fixed number of UAVs, the increase of the sum-payoff signifies that the average individual payoff per UAV increases as a larger number of MNOs is employed. Consequently, the corresponding UAVs will have better QoE. Moreover, the evaluation shows that the proposed solution outperforms the random selection scheme by yielding a larger sum-payoff. Note as well that the gain in terms of sum-payoff obtained by using our proposed solution instead of the random selection increases as we increase the number of MNOs.

Meanwhile, Fig. 4 depicts the number of transfer operations executed by the algorithm before reaching the stability. This reflects the convergence speed of the algorithm. A larger number of transfer operations induces a longer time for the algorithm to reach an optimal stable partition. From Fig. 4, we see that the number of transfer operations increases, in general, with the number of considered MNOs and the connected UAVs. Indeed, these two parameters reflect respectively the number of coalitions and the number of players. The number of players' transfer attempts is executed according to the size of these two parameters (lines 3 and 4 in Algorithm 1). This demonstrates the impact of these two parameters on the convergence speed of the algorithm. On the other hand, we also take note that in few situations, the number of transfer operations can decrease when passing to more MNOs or UAVs. For example, as it can be seen from Fig. 4, the number of transfers considering three MNOs and 60 UAVs is less than that using 55 UAVs. This is due to fact that the initial partition is random (random selection of the serving MNO). As expressed by equation (18), the initial partition is subject to a sequence of player transfer operations until reaching the stability. Each operation allows enhancing the sum of the characteristic function of the coalitions. This shows that the initial partition plays also a role in increasing the convergence speed of the algorithm. If the initial random partition is closer to the final stable partition, a smaller number of transfer operations is needed for the algorithm to converge to the final optimal partition. It is important to mention that the results in Fig. 4 were obtained by averaging over 9 trials. By averaging over several trials, we decrease the variance of the obtained results.

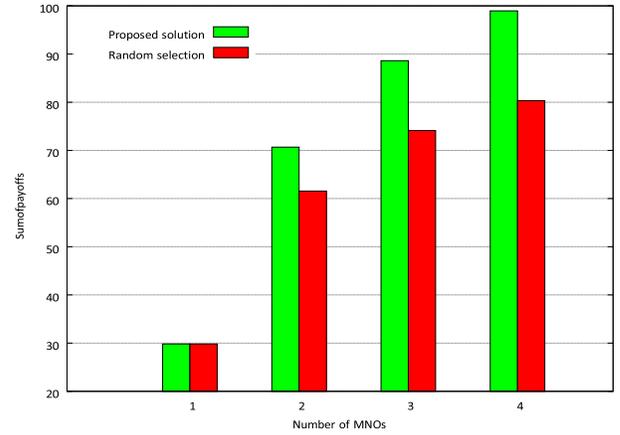

Fig. 3: Evaluation of the sum of payoffs (120 UAVs).

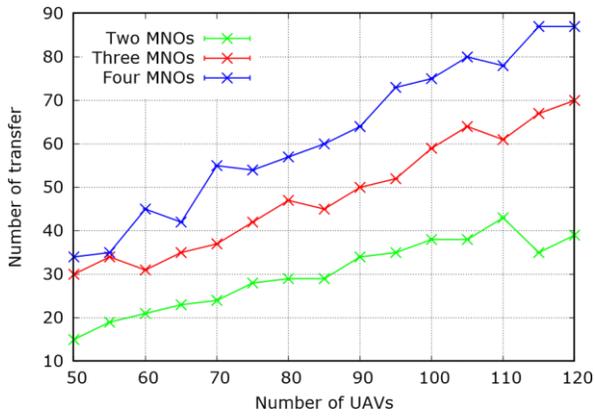

Fig. 4: Evaluation of the number of transfer operations.

## VI. Conclusion

In this paper, a mechanism for connection steering in mobile network-enabled UAVs has been proposed. It aims to select, for each UAV, the MNO that provides the best QoE for transmitting data. To this end, the paper considers a realistic communication model. Given the complexity of the related optimization problem, a coalitional game optimization based solution has been proposed. The goal is to form UAV coalitions around the MNOs in a way to enhance their QoE. A transfer operation has been defined to enable UAVs to change their coalitions, reduce their outage probability, and increase their payoff. Through simulation, we have shown the potential of exploiting several MNOs to enhance the UAVs' QoE. We demonstrated the effectiveness of our proposed coalitional game approach in converging to a stable partition that maximizes the sum-payoff of the aggregate network.

## Acknowledgment

This work was partially supported by the European Unions Horizon 2020 Research and Innovation Programme under the 5G!Drones project (Grant No. 857031) and the EU/KR PriMO-5G project (Grant No. 815191). It was also supported in part by the Academy of Finland 6Genesis project (Grant No. 318927) and CSN project (Grant No. 311654).

## References

[1] 3GPP, "Study on enhanced LTE support for aerial vehicles," *Technical Report,3GPP TR 36.777*, 2017.

[2] H. Hantouti, N. Benamar, T. Taleb, and A. Laghrissi, "Traffic steering for service function chaining," *IEEE Communications Surveys Tutorials*, pp. 1–1, 2018.

[3] T. Taleb and A. Ksentini, "Vecos: A vehicular connection steering protocol," *IEEE Transactions on Vehicular Technology*, vol. 64, no. 3, pp. 1171–1187, March 2015.

[4] B. V. Der Bergh, A. Chiumento, and S. Pollin, "Lte in the sky: trading off propagation benefits with interference costs for aerial nodes," *IEEE Communications Magazine*, vol. 54, no. 5, pp. 44–50, May 2016.

[5] H. Hellaoui, O. Bekkouche, M. Bagaa, and T. Taleb, "Aerial control system for spectrum efficiency in uav-to-cellular communications," *IEEE Communications Magazine*, vol. 56, no. 10, pp. 108–113, OCTOBER 2018.

[6] H. Hellaoui, A. Chelli, M. Bagaa, T. Taleb, and P. Matthias, "Towards Efficient Control of Mobile Network-Enabled UAVs," in *2019 IEEE Wireless Communications and Networking Conference (WCNC 2019)*, April. 2019.

[7] M. Mozaffari, W. Saad, M. Bennis, and M. Debbah, "Efficient deployment of multiple unmanned aerial vehicles for optimal wireless coverage," *IEEE Communications Letters*, vol. 20, no. 8, pp. 1647–1650, Aug. 2016.

[8] N. H. Motlagh, M. Bagaa, T. Taleb, and J. Song, "Connection steering mechanism between mobile networks for reliable uav's iot platform," in *2017 IEEE International Conference on Communications (ICC)*, May 2017, pp. 1–6.

[9] A. Chelli, H. Tembine, and M. Alouini, "A coalition formation game for transmitter cooperation in ofdma uplink communications," in *2014 IEEE Global Communications Conference*, Dec 2014, pp. 4197–4202.

[10] M. Abramowitz and I. A. Stegun, *Handbook of mathematical functions: with formulas, graphs, and mathematical tables*. Courier Corporation, 1964, vol. 55.

[11] H. Hellaoui, A. Chelli, M. Bagaa, and T. Taleb, "Towards Mitigating the Impact of UAVs on Cellular Communications," in *2018 IEEE Global Communications Conference (GLOBECOM 2018)*, Dec. 2018.

[12] A. F. Molisch, *Wireless Communications*. Chichester: John Wiley & Sons, 2005.